 \documentclass{camera}
\usepackage{epsfig}
 \def\ino{\widetilde}
 \def\slepton{\ino{\ell}}

 \def\gsim{\:\raisebox{-0.5ex}{$\stackrel{\textstyle>}{\sim}$}\:}
 \def \ee  {e^{+}e^{-}}
 
 \def \grav {\tilde{{G}}}
\newcommand{\Pcha}{\ensuremath{\chi^{\pm}}}

\newcommand{\PChiz}[1]{\ensuremath{\chi^0_{#1}}}
\newcommand{\PSnu}{\ensuremath{\widetilde{\nu}}}

\newcommand{\PStaur}{\ensuremath{{\tilde{\tau}}_\mathrm{R}}}

\newcommand{\PSe}{\ensuremath{\widetilde{{e}}}}

\newcommand{\PSb}{\ensuremath{{\widetilde{b}}}}

\newcommand{\PSbl}{\ensuremath{{\tilde{b}}_\mathrm{L}}}
\newcommand{\PSt}{\ensuremath{{\widetilde{t}}}}
\newcommand{\PSq}{\ensuremath{{\widetilde{q}}}}

\newcommand{\Pchi}{\ensuremath{\chi}}

\newcommand{\neui}{\ensuremath{\chi^0_{i}}}
\newcommand{\neuj}{\ensuremath{\chi^0_{j}}}

\newcommand{\neuu}{\ensuremath{\chi^0_{1}}}
\newcommand{\neud}{\ensuremath{\chi^0_{2}}}
\newcommand{\neut}{\ensuremath{\chi^0_{3}}}
\newcommand{\neuq}{\ensuremath{\chi^0_{4}}}

\begin{document}

\begin{flushright}
{\large CERN-TH/2001-152}\\
{\large June 2001}\\
\end{flushright}

%
\title{
~~~~ \\
\vspace{-3.0truecm}
SEARCHES FOR NEW PARTICLES AT LEP:\\
A SUMMARY REPORT}

%
%
\author{~~~~\\
\vskip-3.0cm
M. Antonelli$^1$ \And S. Moretti$^2$}

\organization{
\vspace{-5.0truecm}
$^1$Laboratori Nazionali di Frascati INFN, I-00044, Frascati, Italy
$^2$Theory Division, CERN, CH-1211 Geneva 23, Switzerland}

\maketitle

\vspace{-4.0truecm}
\abstract{
\small\it We review the progress made at LEP in the quest
for new particles.}

\vspace*{0.5truecm}
\centerline{\sl Summary talk presented at:}
\centerline{\sl `LEPTRE: XIII Convegno sulla Fisica al LEP',} 
\centerline{\sl Rome, Italy, 18-20 April 2001.}
\vspace*{0.5truecm}

%
\section{Introduction}
\label{Sect:Intro}
Twelve exciting years of research at the high energy frontier
are the legacy of the Large Electron-Positron (LEP) collider
 at CERN. During its runs at the
centre-of-mass (CM) energies of $\sqrt s={M_{Z^0}}$ (LEP1) 
and 130--209 GeV (LEP2), 
this machine has allowed for the collection
of an unprecedented amount of data. About 1 fb$^{-1}$ of integrated 
luminosity has been delivered per experiment.
 Never before as during the LEP
era the Standard Model (SM) of particle physics has
undergone such a stringent, yet so successful, scrutiny of its most 
fine details. 
LEP has now been turned off, and it is our aim here to provide
a comprehensive, yet brief, summary on the subject of searches
for new particles at the CERN collider, both within and beyond the SM, and 
review the prospects at future accelerators.

\section{The SM Higgs boson: the LEP excess} 
\label{Sect:SM-Higgs}
Contrary to what hoped for initially, expectations of LEP clarifying the
mechanism of Electroweak Symmetry Breaking (EWSB)  have 
faded away. 
Within the SM of particle physics, EWSB is realised through
the so-called Higgs mechanism, whose unmistakable hallmark
would be the discovery of a neutral scalar particle, the Higgs boson
(hereafter denoted by $H$).
In the year 2000, LEP operations have been optimised towards the 
SM Higgs boson search
\cite{Ferro-Kado}. As a result of this, the LEP-combined sensitivity for a
3$\sigma$ observation reached 115 GeV, assuming Higgs production 
in association with a $Z^0$ boson, via $e^+ e^- \to {H Z^0}$.
A 115 GeV SM Higgs boson predominantly decays 
into ${b\bar{b}}$ (74\%) and $\tau^+\tau^-$ (7\%).
The analyses addressed the following final states: 
four-jets (${H q\bar{q}}$),
missing energy  (${H \nu\bar{\nu}}$), lepton pairs
(${H \ell^+ \ell^-}$, $\ell=e,\mu$)
 and $\tau$'s  (${H \tau^+ \tau^-}$ plus 
 ${H \to \tau^+ \tau^-, Z^0 \to q\bar{q}}$).

\begin{figure}[ht]
\begin{center}
\epsfig{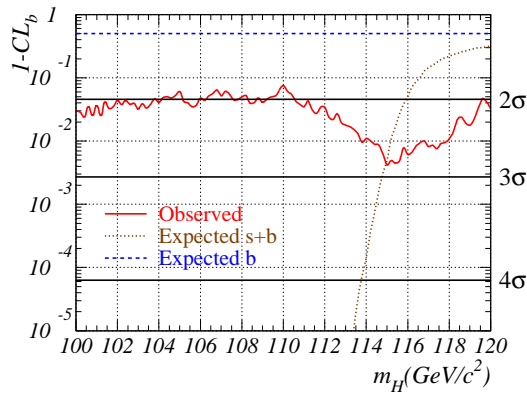} 
\end{center}
\vskip-0.5cm
\caption{\small Confidence level 1-CL$_{\rm{b}}$ (full curve) as a function of 
${M_{H}}$. The dashed (dotted) curve indicates the expected level in the
 background only (signal-plus-background) hypothesis.
 The straight full lines indicate the level 
 2, 3 and 4$\sigma$ excesses above the expected background rate.}
\label{hclb}
\end{figure}

The results of the LEP-combined data presented at the LEPC meeting on November
 3$^{\mathrm{th}}$, 2000, showed 
 an excess of 2.9$\sigma$ beyond the background 
 expectation.  The compatibility of the data with the background-only
 hypothesis can be parametrised as $1-{\rm CL_b}$, see fig.
 \ref{hclb}, as a function of the Higgs mass.
 The distribution exhibits a minimum at 115 GeV. The
 probability that this minimum arises from a background 
 fluctuation is 0.4\%.

 The log-likelihood ratio (LLR) values at ${ M_{H}}$= 115 GeV
 observed in each experiment are such that the result of ALEPH
 is a little too signal-like, DELPHI is more background-like whereas
 L3 and OPAL are close to the most likely value expected for 
 ${M_{H}}$= 115 GeV.
 The distribution of the four observed values is consistent
 with the one expected in the signal-plus-background hypothesis.
 
 In the combined results, in each
 of the four channels, the LLRs observed are
 close to the most likely value expected for 
 ${ M_{H}}$= 115 GeV. The significance of the observed excess 
 is largest in the four-jet channel, followed by the missing energy, 
 the leptonic and the tau one,
 as expected from the decreasing signal-to-background separation for 
 these final states.
 Moreover, the observed significance obtained with the 
 data samples analysed shows a progressive and regular increase
 indicating that the observed effect does not result
 from an early statistical fluctuation, which would have then been 
 reduced by additional statistics.
 
 The observed excess is compatible
 with a Higgs boson with mass near 115 GeV.
 More data, or results from other experiments,
 will be needed to determine whether the observed excess is 
 real. Unquestionably though, if the `same' 115 GeV 
 Higgs boson will eventually be detected, at either the
 Tevatron (Run 2) at FNAL or the Large Hadron Collider (LHC) at
 CERN, much of the credit for its
 discovery will have to remain among LEP  
 achievements.

 In the meantime, a lower limit of 113.5 GeV at 95\% C.L.
 on the SM Higgs mass has been derived,
 {\it i.e.}, about 2 GeV below the median expected 115.3 GeV.

\section{Supersymmetry}
\label{Sect:SUSY}
Despite its innumerable experimental successes, the SM cannot be
a fundamental theory valid up to $M_{\mathrm{Planck}}\sim
10^{18}$ GeV (where a description which includes quantum gravity is needed).
The SM  has to be replaced at an energy higher than the Fermi scale,
$G_F^{-1/2}\approx300$ GeV, by some more fundamental theory.
This can be seen from the fact that the
one-loop radiative corrections to the SM Higgs mass $M_{H}$ are
quadratically divergent
(naturalness or hierarchy problem) \cite{naturalness}.

Supersymmetry (SUSY) can solve the hierarchy problem.
This is a possible symmetry of nature that relates all the SM
fundamental fields (those describing quarks, leptons, gauge and Higgs
bosons) to a new set of physical states (`sparticles'), 
identical to the latter in
everything, except for their spins, which differ by half unit. 
As a consequence of their different statistics, 
cancellations occur between the bosonic and fermionic 
loop contributions to the Higgs mass, 
ensuring that SUSY is free from quadratic divergences. 
SUSY must be broken though,
since we do not observe the `Superpartners' with the same mass
as ordinary matter. 
However, if SUSY has to remain an (approximate)
symmetry above the TeV scale, it must be broken `softly': {\it i.e.},
by terms that do not re-introduce quadratic divergences (rather, only
logarithmic). These soft parameters are dimensionful and, in order not
to contradict naturalness, their mass scale is expected to fall in the TeV
region. Are precisely the values of 
these terms that set the upper scale of
the sparticle masses.
There are various mechanisms of soft SUSY breaking. These have
been reviewed in Ref.~\cite{Rattazzi}, with particular emphasis
on Minimal-Supergravity (MSUGRA) \cite{MSUGRA}
and Gauge Mediated Symmetry Breaking 
(GMSB) \cite{GMSB}, whose signatures have been of particular concern at LEP
\cite{Ranieri-Azzurri}.

\begin{table}
\begin{center}
\small
\begin{tabular}{|lcclcc|}     \hline
Particle & & Spin & Sparticle & & Spin \\
\hline
quark  & $q$ & 1/2 & squarks & $\ino{q}_{L,R}$ & 0 \\
charged lepton & $\ell^\pm$ & 1/2 & charged sleptons 
& $\ino{\ell}_{L,R}^\pm$ & 0 
\\
neutrino  & $\nu$ & 1/2 &  sneutrino & $\ino{\nu}$ & 0 \\
gluon & $g$ & 1 & gluino & $\ino{g}$ & 1/2 \\
photon & $\gamma$ & 1 & photino & $\ino{\gamma}$ & 1/2 \\
neutral gauge boson &$Z^0$ & 1 & zino & $\ino{Z}$ & 1/2 \\
neutral Higgs bosons & $h^0,H^0,A^0$ & 0 & neutral Higgsinos &
$\ino{H}^{0}_{1,2}$ & 1/2 \\ 
charged gauge boson & $W^\pm$ & 1 & wino & $\ino{W}^\pm$ & 1/2 \\
charged Higgs boson & $H^\pm$ & 0 & charged Higgsino &
$\ino{H}^{\pm}$ & 1/2 \\ 
graviton & $G$ & 2 & gravitino/goldstino & $\ino{G}$ & 3/2 \\
\hline
\multicolumn{6}{|c|}{
$\ino{W}^\pm, \ino{H}^\pm$ mix to form 2 chargino 
mass eigenstates $\chi^{\pm}_{1}, \chi^{\pm}_{2}$}\\
\multicolumn{6}{|c|}{
$\ino{\gamma}$, $\ino{Z}$, $\ino{H}^0_{1,2}$ mix to
form 4
neutralino mass eigenstates $\neuu,\neud,\neut,\neuq$}\\ 
\multicolumn{6}{|c|}{${\ino t}_L,{\ino t}_R$ (and similarly
${\ino b}, {\ino\tau}$) mix to form the mass eigenstates
${\ino t}_1, {\ino t}_2$} \\ 
\hline
\end{tabular}
\end{center}
\caption{\small Particle content of the MSSM, expressed in terms of its
mass eigenstates.}
\label{Tab:content}
\end{table}

\subsection{The MSSM}
\label{Sect:MSSM}
M-SUGRA and GMSB scenarios can be accommodated within
the Minimal Supersymmetric Standard Model (MSSM),
wherein the particle content and number of free parameters entering the
Lagrangian is kept to the minimum compatible with SUSY. 
The entire (s)particle spectrum of the MSSM is specified in
table \ref{Tab:content}. 

\subsubsection{The MSSM Higgs bosons}
\label{Sect:MSSM-Higgses}
But let us turn again to the Higgs sector, albeit in the new model. 
A pre-requisite for the realisation of the MSSM is the 
primordial existence of two Higgs doublets, one coupling
to up- and one to down-type (s)fermions (in contrast to the only singlet
field of the SM, which is
 universally coupled), whose non-zero vacuum expectation
values (VEVs) induce spontaneous EWSB. Of the
initial eight degrees of freedom (in a four-dimensional space) of the two
complex Higgs doublet fields of the MSSM, upon EWSB and mass generation in the
gauge boson sector, three are absorbed by the standard weak fields ($Z^0$ and
$W^\pm$) in the form
of a longitudinal polarisation; five instead survive as physical
Higgs states. Of these, three are neutral and two charged. Whereas the
latter have a mixed CP-nature, the former comprehend two CP-even (or scalar)
states, denoted by $h^0$ and $H^0$, and a CP-odd 
(or pseudoscalar) one, labelled as $A^0$.

At tree level, all masses and couplings in the Higgs sector can be
expressed in terms of only two real parameters, the ratio of VEVs
of the two Higgs doublets (denoted by $\tan\beta$)
and the mass of one of the bosons ({\it e.g.}, $M_{A^0}$). In addition,
at lowest order,
one has: $M_{h^0}\le M_{Z^0}$ (see, {\it e.g.}, \cite{HHG}).
However, this upper value on the lightest Higgs boson mass
is significantly modified by virtual effects. At
two-loop order \cite{Higgs-two-loop}, it becomes 130 GeV or
so, largely within the reach of LEP. 
Hence, it is not surprising that most of the efforts spared at the CERN
$e^+e^-$ collider
in detecting physics beyond the SM have actually coincided with the
search for this particular Higgs state. 

\vskip0.20cm\noindent
\underline{\sl Searches for neutral Higgs bosons.~}
 In the MSSM, neutral Higgs bosons are produced  via 
 the Higgsstrahlung process $e^+ e^- \to {h^0Z^0}$ and through
 pair production $e^+ e^- \to {h^0A^0}$.
 The cross section of the former is proportional to 
 $\sin^2(\beta-\alpha)$; the one of the latter to $\cos^2(\beta-\alpha)$.
 In the mass range of interest for LEP2 searches, the main $h^0$ and $A^0$
 decay modes are in ${b\bar{b}}$ and $\tau^+ \tau^-$.
 To search for Higgsstrahlung production the same selections developed
 for the SM signals were used whereas
 two more additional signatures arise
 via $h^0A^0$ production: the `four $b$-jet'
 and the `two $b$-jet plus two tau' final states.
 Good agreement 
 with the expectations from SM processes has been found
 for both topologies and upper limits on the $h^0A^0$ cross section 
 have been set as a function of the $A^0$ and $h^0$ 
 masses (see \cite{Cucciarelli} 
 for more details).
 The results obtained in the searches for $h^0Z^0$ and $h^0A^0$ production
 are interpreted within two scenarios: maximal and minimal mixing in the
 stop sector.  Lower limits on the masses of $h^0$ and 
 $A^0$ have been set:
 ${M_{h^0}}>91$ GeV 
 and  ${M_{A^0}}>92$ GeV for 
 any value of $\tan \beta$.
 In the conservative maximal mixing scenario, the $\tan \beta$ 
 region [0.48-2.56] is excluded at 95\% C.L.

\vskip0.20cm\noindent
\underline{\sl Searches for charged Higgs bosons.~}
 Pair production of charged Higgs bosons occurs mainly via $s$-channel
 exchange of a photon or a $Z^0$ boson. The $H^+$ decays  predominantly
 into $ c \bar{s}$ or $\tau^+ \nu_\tau$ (and 
 charge conjugates). Three final states have been studied at  
 LEP: ${c\bar{s} s \bar{c}}$,
 ${c\bar{s} \tau^- \bar{\nu_\tau}}$/${s\bar{c} \tau^+ \nu_\tau}$
 and $\tau^- \bar{\nu_\tau} \tau^+ \nu_\tau $.
 A $3\sigma$ deviation with respect to the SM background expectation has
 been observed by L3 in the ${c\bar{s} s \bar{c}}$ channel
 for $M_{H^\pm}\approx67$ GeV.
 This effect has not been confirmed by the other experiments 
 (see \cite{Garcia-Abia} 
 for more details).
 A lower limit on $M_{H^\pm}$ has been set at 78 GeV,
 independently of the branching ratio BR($H^+ \to \tau^+ \nu_\tau$).
 The above numbers are valid within a general Two-Higgs Doublet
 Model (2HDM) and may also by applied to the MSSM in 
 the mentioned $\tan\beta$ region, where an indirect lower limit on
 $M_{H^\pm}$ is derived from the one on $M_{h^0}$: 
 $M_{H^\pm}^2\approx M_{W^\pm}^2+M_{h^0}^2\gsim (130~\mathrm{GeV})^2$.

\subsubsection{The MSSM sparticles}
\label{Sect:MSSM-sparticles}
Many more physical states are however expected in the SUSY theory, 
{\it i.e.}, 
the Supersymmetric partners of ordinary matter:
 namely, sleptons, squarks and gauginos/Higgs\-inos
(see table \ref{Tab:content}). Depending upon the mechanism of SUSY
breaking, several signatures involving MSSM sparticles (including
the gravitino) were within the reach of LEP.
(R-parity is assumed to be conserved throughout.) 
\renewcommand{\arraystretch}{1.1}
\begin{table}[t]
 \begin{center}
 \begin{tabular}{|l|l|l|}
  \hline
Production  & Decay mode & Topology  \\
\hline
\hline
 $\slepton \bar{\slepton}$ & $\slepton \to \ell \chi^0_1$ & Acoplanar leptons   \\
 $\PSe_{\rm{L(R)}} \bar{\PSe}_{\rm{R(L)}}$ & $\PSe \to {e} \chi^0_1$ &
 Single electron  (small $m_{\PSe_{\rm{R}}}\!-\!m_{\chi^0_1}$) \\
\hline\hline
\rule{-4pt}{14pt} $\PSq\bar{\PSq}$ &
 $\PSq \to {q} \chi^0_1$ & Acoplanar jets \\
 $\PSt\bar{\PSt}$ &
 $\PSt \to {c} \chi^0_1$ & Acoplanar jets \\
$\PSb\bar{\PSb}$ &
 $\PSb \to {b} \chi^0_1$ & Acoplanar $b$-jets \\
 $\PSt \bar{\PSt}$ & $\PSt \to {b} \ell \PSnu $ &
Acoplanar jets plus leptons \\
\hline\hline
$\Pchi^+\Pchi^-$ & $\Pcha\!\to\!{q \bar{q}^\prime}\chi^0_1$ & 4 jets + $\not{\!\!E}$  \\
           & $\Pcha\!\to\!\ell^\pm \nu\chi^0_1$ & Acoplanar leptons  \\
           & mixed & 2 jets + lepton + $\not{\!\!E}$  \\
\hline\hline
\neui\neuj         & $\neuu\neuj\!\to\!{q \bar{q}}\neuu$ & 
                            \raisebox{-3mm}[0mm][-3mm]{Acoplanar jets} \\
                   & $\neui\neuj\!\to\!\nu\bar{\nu}\neuu 
                      {q \bar{q}}\neuu,\ldots$ & \\
$j\!\geq\!i, j\!\neq\!1$ & $\neuu\neuj\!\to\!\ell^+\ell^-\neuu$ 
                            & \raisebox{-3mm}[0mm][-3mm]{Acoplanar leptons} \\
                         & $\neui\neuj\to\!\nu\bar{\nu}\neuu \ell^+\ell^-\neuu,\ldots$ 
                                                        &  \\
\hline
\end{tabular}
\end{center}
\caption{\small Final state topologies studied in  MSUGRA.}
\label{topologies_mssm}
\end{table}
\renewcommand{\arraystretch}{1.0}
%
\vskip0.20cm\noindent
\underline{\sl Searches for MSUGRA topologies.~}
 In the MSUGRA scenario
 the Lightest Supersymmetric Particle (LSP) is the lightest neutralino
 ($\chi^0_1$)
 and the gravitino is heavier than the other SUSY particles.
 The final states topologies addressed by MSUGRA searches are summarised 
 in table \ref{topologies_mssm}
 (only the main decay chains contributing to the different topologies 
  are indicated for neutralinos). For a given final state, 
 various selection criteria are applied, which depend mainly on 
 the mass difference $\Delta M$ between the produced sparticle 
 and the LSP.
 The number of events selected by the analyses is in good agreement 
 with the expectation from SM processes. A slight excess
 was observed in the acoplanar $\tau$-search in the 1998 and 1999 data. It has
 not been confirmed by the analysis of the 2000 data sample.
 Lower limits on slepton and squark masses are given in 
 table \ref{limits_mssm}.

\renewcommand{\arraystretch}{1.2}
\begin{table}[ht]
 \begin{center}
 \begin{tabular}{|l|c|c||l|c|c|}
\hline
 Particle         & Limit & Conditions of validity  &
 Particle         & Limit & Conditions of validity  \\
\hline
 selectron & 99 & $\Delta M>\!10$~ &
stop      & 92 & $\PSt\!\to\!{c}\PChiz{1}$,
                               $6 < \Delta M < 40$ \\ 

\hline
 smuon     & 95 & $\Delta M>\!10$~, 
                   \mbox{$\tilde{\mu}\!\to\!\mu\PChiz{1}$} &
stop                  & 93 & $\PSt\!\rightarrow\!{b}
         \ell\widetilde{\nu}$,
                       $\Delta M > 10$ \\
\hline
 stau      & 80 & $\Delta M>\!10$~, 
                   \mbox{$\tilde{\tau}\!\to\!\tau\PChiz{1}$}, 
                   \PStaur  &
sbottom   & 96 & $\PSb\!\to\!{b}\PChiz{1}$, $\Delta M > 8$,
                       \PSbl \\
\hline
\end{tabular}
\end{center}
\caption{ \small Lower limits at 95\% C.L. on squark and slepton masses
in MSUGRA. For sleptons, $\tan\beta=2$, $\mu=-200$ GeV. 
All masses and mass differences are in GeV.
}
\label{limits_mssm}
\end{table}
\renewcommand{\arraystretch}{1.}

 Chargino pair production and neutralino associated production are excluded 
 up to the kinematic limit over a significant fraction of the MSSM  
 parameter space.
 From the negative outcome of chargino and neutralino searches, 
 a lower limit on the lightest neutralino mass can be derived as a
 function of $\tan \beta$ for a large scalar mass $m_0$ ({\it i.e.},
 the mass common to all SUSY scalar states at the unification scale).
 The loss of sensitivity of chargino and neutralino searches at 
 low $m_0$ is recovered through slepton searches. A scan performed over
 the relevant parameter space shows that a lower limit on the 
 LSP mass of 38 GeV holds for all values of $m_0$.
 
 Constraints from Higgs boson searches are also used to
 improve the lower limit on the LSP mass.
 As expected, the limit is strongest for low values of
 $\tan \beta$ and $m_0$. The lower limit on the LSP is
 45 GeV for a top mass of 175 GeV.
 These results are however quite 
 sensitive to $m_t$: {\it e.g.}, the limit
 becomes 40 GeV for a top mass of 180 GeV.
 Besides,
 the interplay among the searches for sleptons, charginos, Higgs 
 bosons and the $Z^0$ width measurement at LEP1 can also be exploited in the
 framework of MSUGRA. This way, the
 lower limit on the LSP mass is close to $\sqrt{s}/4$, 
 half the lower limit on the chargino mass.

\vskip0.20cm\noindent
\underline{\sl Searches for GMSB topologies.~}
In GMSB scenarios the LSP
is the weakly-coupled gravitino ($\grav$). Hence,
 in $\ee$ collisions, SUSY sparticles typically
 decay to their SM partner plus gravitinos. 
The  Next-to-LSP (NLSP) is here, in general, 
either the lightest neutralino or 
slepton ({\it e.g.},
three degenerate NLSPs or the stau if its mixing is large). These are
expected to be much lighter than the other SUSY sparticles 
and therefore the only ones accessible at LEP.

The lifetime of the NLSP depends on the gravitino mass (or, equivalently on
the SUSY-breaking scale $\sqrt{F}$). For quite heavy
gravitinos the decay length associated to the lifetime can be comparable to or 
even larger than the size of the LEP detectors. For such a reason, 
topological searches enabling to identify a long-lived or even 
stable NLSP have been developed. A partial list of experimental topologies 
considered is given in table \ref{topologies_gmsb}.

\begin{table}[ht]
{\small
\begin{center}
\vspace{0.4cm}
\begin{tabular}{|l|l|l|l|l|}
\hline
 NLSP & Production & Decay mode & NLSP Lifetime & Exp. Topology  \\
\hline
\hline
 $\chi^0_1$  & $\ee \to \chi^0_1 \chi^0_1$ & $\chi^0_1 \to \gamma \grav$  &  $c \tau << \ell_{\rm{detector}}$ &
 Acoplanar $\gamma$'s  \\
 &  &  &  $c \tau \sim \ell_{\rm{detector}}$ & Non Pointing $\gamma$  \\
 &  &  &  $c \tau >> \ell_{\rm{detector}}$ & Invisible  \\
\hline
\hline
 $\slepton$  & $\ee \to \slepton \slepton$ & $\slepton \to \ell \grav$  &
  $c \tau << \ell_{\rm{detector}}$ & Acoplanar $\ell$'s  \\
 &  &  &  $c \tau \sim \ell_{\rm{detector}}$ &  Large impact parameter tracks  \\
 &  &  &  $c \tau >> \ell_{\rm{detector}}$ & Heavy Stable Charged Particles   \\
\hline
 $\slepton$  & $\ee \to \chi^0_1 \chi^0_1$ & $\chi^0_1 \to \slepton \ell \to \ell \ell \grav$  &
  $c \tau << \ell_{\rm{detector}}$ & Multi-$\ell$'s (2 hard and 2 soft)  \\
 &  &  &  $c \tau \sim \ell_{\rm{detector}}$ & Not yet studied  \\
 &  &  &  $c \tau >> \ell_{\rm{detector}}$ & Not yet studied  \\

\hline
\end{tabular}
\end{center}
}
\caption{\small Final state topologies studied in GMSB.}
\label{topologies_gmsb}
\end{table}

 No evidence for any such processes has been found in the data
 and lower limits on the sparticle masses have been set.
 The stau is excluded up to a mass of 80 GeV for any
 lifetime. In the case of neutralino NLSP the limit depends
 strongly on the neutralino lifetime.
 For lifetimes short enough the neutralino decays 
 via $\chi^0_1 \to \gamma \grav$ and can be detected
 directly. The searches for acoplanar photons and non pointing single-photons
 \cite{DeMin}
 set a lower limit on $M_{\chi^0_1}$ of about 70 GeV for
 $c\tau$ up to 10 m.
 For longer lifetimes only indirect searches can be used and
 the lower limit on the neutralino mass is close to the one obtained in
 the MSUGRA model.

 It has been shown that photonic final states
 can probe theories with extra spatial dimensions \cite{DeMin}. Here,
 one expects additional contributions to $e^+ e^- \to \gamma \gamma$ 
 due to virtual graviton exchange as well as direct production of the
 latter via $e^+ e^- \to \gamma G$.   
Single-photon final states naturally accommodate also signatures
induced by light, so-called, `sgoldstinos' \cite{Perazzi}\footnote{In fact,
in Supersymmetric extensions of the SM with a very light gravitino, 
the effective theory at the EW scale should contain not only the
Goldstino, but also its partners from SUSY, the sgoldstinos 
(two neutral spin-less sparticles).}. 

\subsubsection{MSSM and dark matter}
\label{Sect:DM}
The extensive searches performed at LEP have ruled out a large
 fraction of the MSSM parameter space interesting for cold dark 
matter \cite{Ganis}. 
 The LEP results are compatible over a small region of the parameter space 
 with cosmological constraints and with the
 SUSY interpretation of the
 disagreement between the expected and measured values of
 the muon anomalous magnetic moment.

\subsection{Beyond the MSSM: MSSM$^*$ and NMSSM.}
\label{Sect:Beyond-MSSM}
The model embodying SUSY need not be minimal. Indeed, several `extensions'
of the MSSM have been considered in literature. By extensions,
we mean here theoretical setups which embed a number of
parameters in the SUSY Lagrangian larger than those appearing in the canonical
MSSM. For example, this can be done by either dismissing
the assumption that the mentioned soft SUSY-breaking terms are
real (hence taking these as complex) \cite{MSSM*} or adding one 
singlet Higgs field (and its SUSY counterpart) 
\cite{NMSSM}\footnote{We do not consider here the
possibility of additional Higgs doublets or triplets: see, {\it e.g.},
Ref.~\cite{HHG}.}.  Hereafter, we denote the first category of models as
MSSM$^*$ and the second as NMSSM (for Next-to-MSSM). 

MSSM$^*$ scenarios rely on cancellations \cite{cancellations} among
SUSY contributions to the electron and 
neutron Electric Dipole Moments (EDMs) \cite{EDMs},
in order to be realistic\footnote{Despite the 
scope for very large phases
has significantly been reduced recently \cite{Steve}, in view of 
the mercury EDM measurement \cite{mercury}, the extended model has gathered
much interest lately \cite{Gordy}.}. NMSSM settings have been introduced
as a possible solution to the so-called $\mu$-problem of the MSSM, {\it i.e.},
the `unnatural' presence of the $\mu\hat{H}_u\hat{H}_d$ term in the soft SUSY
Lagrangian ($\hat{H}_{u,d}$ are the Higgs(ino) Superfields).

\subsubsection{MSSM$^*$ and NMSSM at LEP and future colliders}
In respect to LEP physics, the effect in either scenario is mainly 
to alter the phenomenology
of the Higgs sector, by inducing a modification to the Higgs masses and
couplings, via
mixing effects
affecting the ordinary neutral Higgs fields of the MSSM, either among each 
other (MSSM$^*$) or with new ones (NMSSM). 
In the MSSM$^*$, this is achieved through one-loop effects 
\cite{Accomando},
induced by explicit CP-violation in the third generation of
squarks, so that 
the three neutral Higgs states are no longer either positive or negative
CP-eigenstates, rather a superposition of the two. In this scenario, {\it
e.g.}, for low and intermediate $\tan\beta$ (say, below 7), the lightest 
Higgs boson mass may be as low as 80--90 GeV and can
have escaped LEP searches because of its reduced couplings to 
$Z^0$ bosons, while the second lightest Higgs mass could be
consistent with the 115 GeV excess (at low $\tan\beta$) \cite{Higgs-mixing}.
In the NMSSM too, the coupling of
the lightest Higgs scalar to gauge bosons can be small, so that,
again, it is  the second lightest Higgs state the observable one 
\cite{Ul}. 
For appropriate combinations of the (reduced) Higgs couplings, the outcome 
here can be the same as in the previous case, with an 
unobservable light Higgs state and the next one lying at 115 GeV.

At future hadronic machines, the phenomenology of either the 
MSSM$^*$ or the NMSSM have not been investigated yet in great detail. 
Only some theoretical studies exists to date and they all focus on the 
Higgs sector: for the MSSM$^*$ see \cite{Sakis} whereas for 
the NMSSM see \cite{Cyril}.

\section{Leptoquarks (LQs)}
\label{Sect:LQ}
These are bosonic fields carrying simultaneously leptonic
and baryonic quantum numbers. They provide a clear signature
of models attempting to explain  the observed 
symmetry between leptons and quarks, with respect to the multiplet structure 
of the EW interactions, such as technicolor \cite{technicolor},
compositeness \cite{compositeness}, Grand Unification Theories
(GUTs) \cite{GUTs} and 
Superstring-inspired scenarios \cite{superstrings}. Evidence of these
 particles has been
 searched for (in vain) at LEP, both directly, {\it i.e.}, via 
 pair and single LQ production, as well as through
$t,u$-channel contributions to, {\it e.g.}, quark pair production
(LQ exchange). 
A review of LEP results and their comparison with the ones
 obtained at Tevatron and Hera is given in \cite{Brigliadori}.

\section{Searches at Tevatron (Run 2) and LHC}
\label{Sect:future}
 The experimental program at Tevatron and LHC is 
largely focused on the
 detection of new physics.
 Detailed studies \cite{Vataga} show that 
 the expected sensitivity at the Tevatron collider in
 direct searches for SUSY particles is slightly beyond the LEP2 
 constraints and not sufficient to guarantee full coverage of the
 SUSY spectrum. 
 In contrast, if SUSY exists at the TeV scale, it is expected
 not to escape  
 experimental detection at the LHC. Besides, 
 it has been pointed out, see \cite{Polesello},  that LHC
 experiments will also be able, in some cases, to
 determine the mechanism of SUSY breaking and the SUSY 
 parameters themselves in various scenarios.

 The reach in discovery of a light neutral Higgs state (in the SM or,
 alternatively, in the MSSM in the low $\tan\beta$ region)
 at the Tevatron collider (Run 2)
 seems to be very promising.
 In particular, it has been shown in \cite{Vataga}
 that a 3$\sigma$ sensitivity for a 115 GeV Higgs mass
 (about the same reached at LEP)
 can be achieved with 3 fb$^{-1}$, corresponding to about two
 years of data taking.
 
Given the strong expectations at both these colliders
concerning the possible detection of a light Higgs boson, it
is of extreme importance the ability to rely on accurate theoretical
predictions. In this respect, it is
worth recalling that important progress has recently
been made in the QCD calculation of the NNLO corrections to
$gg\to $ Higgs \cite{Grazzini}. 
Similarly, one should expect also the QCD NLO
corrections to $q\bar q,gg\to Q\bar Q$ Higgs (with $Q=b,t$)
to become available soon\footnote{See 
Ref.~\cite{WJSZK} for a review of the status of the other
two main Higgs production channels in hadron-hadron collisions: 
$q\bar q^{(')}\to V$ Higgs and
$q\bar q^{(')}\to q^{('')}\bar q^{(''')}$ Higgs (via
$VV$-fusion), with $V=W^\pm,Z^0$.}.

\vskip0.25cm\noindent
{\bf Acknowledgements:}
the authors would like to thank the organisers of the conference
for the friendly and stimulating atmosphere. 

\end{document}